\documentclass[11pt]{article}
\usepackage{amsfonts}

\usepackage{amssymb}
\usepackage{latexsym}
\usepackage{graphicx}
\usepackage[english]{babel}

\begin{document}
\input epsf

\makeatletter
\@addtoreset{equation}{section}
\makeatother


 \begin{center}
{\LARGE Unwinding of strings thrown into a fuzzball}
\\
\vspace{18mm}
{\bf   Stefano Giusto$^\Diamond$\footnote{stefano.giusto@cea.fr} and Samir D. Mathur$^\sharp$\footnote{mathur@mps.ohio-state.edu}
\\}
\vspace{8mm}

$\Diamond$
Laboratoire de Physique Th\'eorique et Hautes Energies\\
Universit\'e Pierre et Marie Curie - Paris 6\\
4 Place Jussieu, 75252 Paris cedex 05, France\\
\vspace{8mm}

$\sharp$ 
Department of Physics,\\ The Ohio State University,\\ Columbus,
OH 43210, USA\\ 
\vspace{10mm}

\end{center}

\thispagestyle{empty}

\def\p{\partial}
\def\h{{1\over 2}}
\def\be{\begin{equation}}
\def\bea{\begin{eqnarray}}
\def\ee{\end{equation}}
\def\eea{\end{eqnarray}}
\def\d{\partial}
\def\la{\lambda}
\def\eps{\epsilon}
\def\bb{\bigskip}
\def\mm{\medskip}
\newcommand{\dm}{\begin{displaymath}}
\newcommand{\edm}{\end{displaymath}}
\renewcommand{\b}{\tilde{B}}
\newcommand{\gm}{\Gamma}
\newcommand{\ac}[2]{\ensuremath{\{ #1, #2 \}}}
\renewcommand{\ell}{l}
\newcommand{\z}{\ell}
\newcommand{\newsection}[1]{\section{#1} \setcounter{equation}{0}}
\def\bb{$\bullet$}
\def\Qbar{{\bar Q}_1}
\def\QPbar{{\bar Q}_p}

\def\q{\quad}

\def\bn{B_\circ}

\let\a=\alpha \let\b=\beta \let\g=\gamma \let\d=\delta \let\e=\epsilon
\let\c=\chi \let\th=\theta  \let\k=\kappa
\let\l=\lambda \let\m=\mu \let\n=\nu \let\x=\xi \let\r=\rho
\let\s=\sigma \let\t=\tau
\let\vp=\varphi \let\vep=\varepsilon
\let\w=\omega      \let\G=\Gamma \let\D=\Delta \let\Th=\Theta
                     \let\P=\Pi \let\S=\Sigma

\def\h{{1\over 2}}
\def\t{\tilde}
\def\r{\rightarrow}
\def\nn{\nonumber\\}
\let\bm=\bibitem
\def\Kt{{\tilde K}}
\def\b{\bigskip}

\let\p=\partial

\begin{abstract}

\b

The traditional black hole has a horizon, with a singularity inside the horizon. But actual microstates of black holes are `fuzzballs', with no horizon and a complex internal structure. We take the simplest hole in string theory -- the extremal 2-charge D1D5 hole -- and study a simple effect that is a consequence of this internal structure of the fuzzball. Suppose we have a NS1 string wrapping the compact circle of the fuzzball solution. In the traditional black hole solution this circle is directly tensored with the remaining directions, and does not shrink to zero size. Thus a part of the string can fall behind the horizon, but not `unwind'. In the fuzzball geometry, this circle makes a nontrivial geometric structure -- the KK monople -- by mixing with the other directions, and thus shrinks to zero at the core of the monopole. Thus the string can `unwind' in the fuzzball geometry, and the winding charge is then manifested by a nontrivial field strength living on the microstate solution. We compute this field strength for a generic microstate, and comment briefly  on the physics suggested by the unwinding process.

\end{abstract}
\vskip 1.0 true in

\newpage
\renewcommand{\theequation}{\arabic{section}.\arabic{equation}}

\def\p{\partial}
\def\r{\rightarrow}
\def\h{{1\over 2}}
\def\b{\bigskip}

\def\nn{\nonumber\\ }

\section{Introduction}
\label{intr}\setcounter{equation}{0}

We have learnt in recent years that the interior structure of black holes is very complex: the information in the hole appears to be distributed over a horizon sized region termed a `fuzzball' \cite{lm4, lm5,lmm,kst,otherpapers}. What are the physical consequences of such a structure? In this paper we take the simplest states of the simplest hole -- coherent states of the 2-charge extremal D1D5 extremal system  -- and study a dynamical process that makes use of the fuzzball structure of the microstate.

We consider type IIB string theory, and compactify spacetime as $M_{9,1}\r M_{4,1}\times S^1\times T^4$. We wrap $n_1$ D1 branes on $S^1$, and $n_5$ D5 branes on $S^1\times T^4$. This system has a microscopic  entropy $S_{micro}\approx 2\sqrt{2}\pi\sqrt{n_1n_5}$ arising from the degeneracy of the extremal D1D5 bound state.

The general microstate from this collection is very quantum, but we will restrict attention to microstates that are described by classical configurations. (This classical limit is described in detail in \cite{fuzzreview1,fuzzreview2}; and we summarize it  in the next section.) These classical microstates are described by geometries that can be explicitly written down. These geometries have no horizons, but they do have an interesting structure. The $S^1$ shrinks to zero along a certain curve ${\cal S}$ in $M_{4,1}$, but in such a way that it does not cause any pathology; the points on ${\cal S}$ become the centers of Kaluza-Klein monopoles, and the full geometry remains non-singular. 

\begin{figure}[ht]
\begin{center}
\includegraphics[width=15cm]{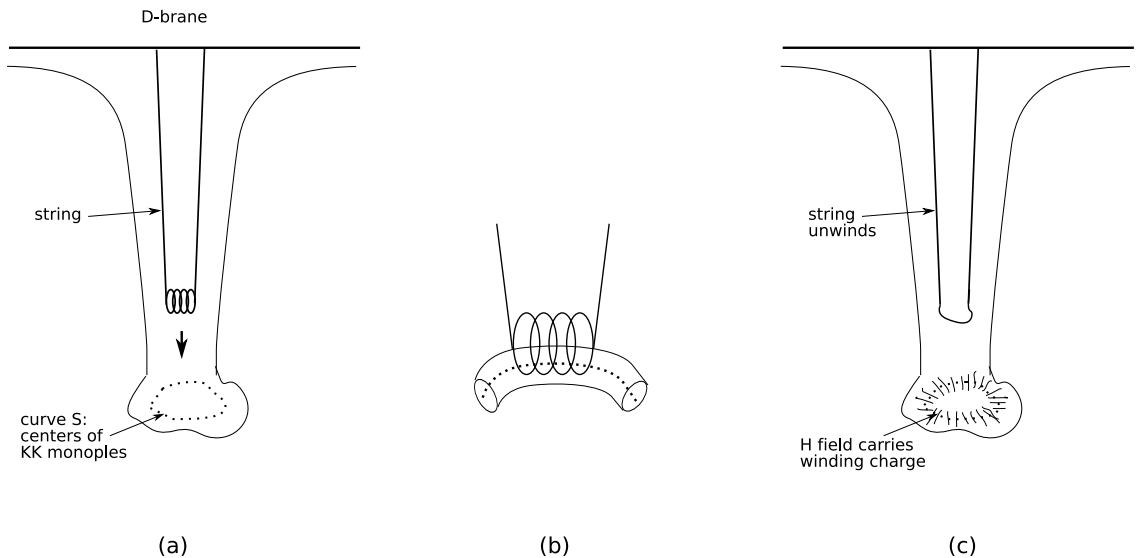}
\caption{(a) A string carrying winding along the $S^1$ is thrown in towards the fuzzball region of a D1D5 microstate (b) Near the curve ${\cal S}$ characterizing the microstate the geometry is that of a KK monopole times a line segment (c) The string can unwind, leaving its winding number as a gauge field excitation on the fuzzball.}\label{faone}
\end{center}
\end{figure}

Now consider an elementary string at $r=\infty$ wrapped along the $S^1$ with winding number $n_\mathrm{NS1}$. This winding charge is a conserved quantum number in this compactification. But if we send the string into the interior of the fuzzball geometry, then the $S^1$ that it wraps will shrink to zero along the points of the curve ${\cal S}$. Thus the string can unwind when it hits this curve, and return to infinity with a winding number different from $n_\mathrm{NS1}$ (fig.\ref{faone}). 

Does this mean that in the presence of the fuzzball we violate conservation of the winding charge? No, because if a string unwinds by reaching the center of a Kaluza-Klein monopole then it generates (in the process of unwinding) a $H^{(3)}$ flux on the monopole, so that the monopole now carries the same winding charge that the string initially had. This was shown for the case of a simple Kaluza-Klein monopole in \cite{ghm}, where an explicit construction of the resulting $H^{(3)}$ field was given. Our goal here will be to show that a similar flux can be placed on the D1D5 microstate geometry, so that as the string unwinds in the geometry we generate a $H^{(3)}$ field to conserve the overall winding charge.

To summarize, we observe that microstate geometries of the D1D5 extremal hole have the following property: an elementary string with winding number along the $S^1$ can interact with the microstate configuration and unwind, and the microstate geometry will pick up the winding charge in the process. 

\section{The D1D5 microstates}\label{solution}
\setcounter{equation}{0}

We take IIB string theory compactified as
\be
M_{9,1}\r M_{4,1}\times S^1\times T^4\,.
\label{qfour}
\ee
We wrap $n_1$ D1 branes on $S^1$ and $n_5$ D5 branes on $S^1\times T^4$. The resulting bound state is called the D1D5 system. By S,T dualities we can make this to the NS1-P system:
\bea
n_5 ~~\mathrm{D5~ branes} ~&\r&~ n_5\equiv n'_1~ \mathrm{NS1~ branes ~(strings)~along ~ S^1}\nonumber \\
n_1 ~~\mathrm{D1~ branes} ~&\r&~ n_1\equiv n'_p~ \mathrm{units~of~momentum ~P~along ~S^1}\,.
\eea
Suppose the length of the $S^1$ is $L'$ after these dualities. We can write the total momentum charge as
\be
P={2\pi n'_p\over L'}={2\pi n'_1n'_p\over n'_1L'}\equiv {2\pi n'_1n'_p\over L_T}\,,
\ee
where $L_T=n'_1 L'$ is the total length of the NS1. The momentum on this multiwound  NS1 comes in units of ${2\pi \over L_T}$, with one quantum of the $k$th harmonic having momentum ${2\pi k\over L_T}$. Thus if we have $N_k$ units of the $k$th harmonic, then we must have
\be
\sum_k k~N_k  =n'_1n'_p=n_1n_5\,.
\label{two}
\ee
The entropy of this extremal system \cite{sen}
\be
S_{micro}=2\sqrt{2}\pi\sqrt{n'_1n'_p}=2\sqrt{2}\pi\sqrt{n_1n_5}
\label{one}
\ee
 arises from the different ways of partitioning the total momentum $P$ among different harmonics. (To get the exact coefficient in this entropy we have to use the fact that there are 8 transverse bosonic oscillation modes and 8 fermionic superpartners of these modes, and the momentum is shared by all these degrees of freedom.)

Each mode of oscillation of the NS1 string can be viewed as a harmonic oscillator. The generic NS1-P state contributing to the entropy (\ref{one}) has excitation number $\sim 1$ for each typical harmonic oscillator, so that the oscillator can be described by a wavefunction, but not a good classical solution. The entire state of the NS1 string is correspondingly very quantum, and we call the generated solution a `fuzzball'. But we can start by looking at simple states where we satisfy (\ref{two}) by  putting a large excitation energy in a few harmonics
\be
N_k\gg 1, ~~~\sum_k k~N_k =n_1n_5\,.
\label{conditionq}
\ee
 Now in each oscillator we excited we can approximate the state by a coherent state, which gives the corresponding mode a classical periodic oscillation. The entire string is now described by a vibration profile $\vec F(y-t)$, where $y$ is the coordinate along the $S^1$, $t$ is the time, and the vector $\vec F$ gives the transverse displacement of the string from its mean position. We can now write the metric produced by this string, and dualize back to the D1D5 duality frame. The microstate thus obtained is given by the solution \cite{lm4}
\bea
&&ds^2 = (H_1 H_5)^{-1/2}[-(dt-A)^2 + (dy+B)^2]+(H_1 H_5)^{1/2} dx_i dx_i + \Bigl({H_1\over H_5}\Bigr)^{1/2} dz_a dz_a\,,\nn
&&F^{(3)}=d[H_1^{-1}(dt-A)\wedge (dy+B)]- *_4 dH_5\,,\nn
&&e^{2\Phi}={H_1\over H_5}\,,
\label{three}
\eea
with
\bea
&&H_5 = 1+{Q_5\over L_T}\int_0^{L_T} {dv\over |x-F(v)|^2}\,,\quad H_1 = 1+{Q_5\over L_T}\int_0^{L_T} dv{|\dot{F}(v)|^2\over |x-F(v)|^2}\,,\nonumber\\
&&A=A_i dx^i  = -{Q_5\over L_T}\int_0^{L_T} dv {\dot{F}_i(v)\over |x-F(v)|^2}\,dx^i\,,\quad dB = -*_4 dA\,.
\label{harmonicfunctions}
\eea
The coordinates $x_i$ ($i=1,\ldots,4$) parameterize the 4 noncompact space directions,  and in these solutions we have restricted the string to vibrate only in these noncompact directions. The Hodge dual on this flat $\mathbb{R}^4$ is denoted as $*_4$; the $T^4$ directions are $z_a$ ($a=1,\ldots,4$). The parameter $Q_5$ measures the D5 brane charge of the geometry. The D1 brane charge is given by
\be
Q_1 = {Q_5\over L_T}\int_0^{L_T} dv |\dot{F}(v)|^2\,.
\ee

The advantage of taking these classical solutions (\ref{conditionq}) (rather than the generic solution) is that we can let an infalling string evolve as a `test particle' on the background, without having to worry about the backreaction caused by the test string. 

\subsection{The curve ${\cal S}$ and the near-curve geometry}\label{near}

The solution (\ref{three}) appears to have a singularity along the curve
\be
\vec x=\vec F(v)\,.
\ee
Let us call this curve ${\cal S}$; it is a simple closed curve in the 4-d noncompact space directions $x_1, \dots x_4$. 
In \cite{lmm} it was shown that the singularity here is only a coordinate singularity; each point along this curve is a center for a Kaluza-Klein monopole. Let us review this fact (we use the notation in the review \cite{fuzzreview1}).

Let us go to a point on the curve given by $v=v_0$. Consider the flat metric $dx_idx_i$ in the four noncompact directions $x_1\dots x_4$. Let $z$ be a coordinate that measures distance along the curve in this flat metric. Still using this flat metric,  choose spherical polar coordinates $(\rho, \theta, \phi)$ for the 3-plane perpendicular to the curve. Then we have
\be
z\approx  |\dot {\vec F}(v_0)| (v-v_0)\,.
\ee
The harmonic functions (\ref{harmonicfunctions}) become
\be
H_5\approx {Q_5\over  L_T}\int_{-\infty}^\infty {dv\over \rho^2+z^2}={Q_5\over  L_T}{1\over |\dot {\vec F}(v_0)|} \int_{-\infty}^\infty {dz\over \rho^2+z^2}={Q_5\pi\over L_T|\dot {\vec F}(v_0)|}{1\over \rho}\,,
\label{hfone}
\ee
\be
H_1\approx {Q_5\pi |\dot {\vec F}(v_0)|\over  L_T}{1\over \rho}\,,\quad
A_z\approx -{Q_5\pi\over L_T}{1\over \rho}\,.
\label{hftwo}
\ee
Let
\be
\tilde Q\equiv {Q_5\pi\over  L_T}\,.
\ee
The relation $dB=-*_4 dA$ gives $B_\phi=-
\tilde Q(1-\cos\theta)$. The $(y, \rho, \theta, \phi)$ part of the metric  becomes
\bea
ds^2&\rightarrow& (H_1 H_5)^{-1/2}(dy+B_i dx^i)^2+(H_1 H_5)^{1/2}(d\rho^2+\rho^2(d\theta^2+\sin^2\theta d\phi^2))\nonumber\\
&\approx&{\rho\over \tilde Q}(dy-{\tilde Q}(1-\cos\theta) d\phi)^2+ {\tilde Q\over \rho}
(d\rho^2+\rho^2(d\theta^2+\sin^2\theta d\phi^2))\,.
\label{vfourt}
\eea
This is the metric near the core of a Kaluza-Klein monopole, and is smooth if the length of the $y$ circle is 
\be
L_y=2\pi R_y=4\pi \tilde Q\,.
\label{relation}
\ee
 It turns out that the relation (\ref{relation}) is exactly satisfied \cite{lmm}. To show this one needs  to construct the NS1-P metric with source as the NS1 string with its correct tension $T={1\over 2\pi\alpha'}$. This tension is related to the   gravitational constant through $G_N=8\pi^6 g^2\alpha'^4$. Dualizing the solution to the D1D5 frame we find that the $S^1$ (parametrized by $y$) has  the correct radius for a smooth Kaluza-Klien monopole geometry, so there is no singularity due to the shrinkage of the $S^1$. The change of coordinates at the KK center
\bea
&&{\tilde r}^2=\rho, ~~
\tilde\theta={\theta\over 2}, ~~\tilde y={y\over 2{\tilde Q}}, ~~\tilde \phi=\phi-{y\over 2{\tilde Q}}
\nonumber\\ 
&&0\le\tilde\theta<{\pi\over 2}, ~~0\le \tilde y<{\pi R_y\over {\tilde Q}}=2\pi, ~~0\le\tilde\phi<2\pi\,,
\label{fifty}
\eea
makes manifest the locally $R^4$ form of the metric
\be
ds^2=4{\tilde Q}[d\tilde r^2+\tilde r^2(d\tilde\theta^2+\cos^2\tilde\theta d\tilde y^2+\sin^2\tilde\theta d\tilde\phi^2)]\,.
\ee

The $(t,z)$ part of the geometry gives
\bea
ds^2&\rightarrow&-(H_1 H_5)^{-1/2}(dt-A_zdz)^2+ (H_1 H_5)^{1/2} dz^2\nonumber\\
&\approx&-{\rho\over \tilde Q}dt^2-2 dtdz\approx -2dtdz\,,
\label{qthree}
\eea
which is regular. The $T^4$ part gives $|\dot {\vec F}(v_0)| dz_a dz_a$ and is thus regular as well. 

\section{Unwinding strings}
We have noted that the D1D5 microstates look, near the curve $\mathcal{S}$, as a  KK monopole tensored with a line segment.  One thus expects the physics of a string thrown into one such microstate to be similar to that in a KK monopole geometry. We first review the physics of string unwinding in a KK monopole geometry, and then discuss the extension of this behavior to our microstate geometries. 
 
\subsection{Unwinding strings in the pure KK solution}\label{unwinding1}

In \cite{ghm} the following interesting problem was studied. Suppose we have the basic KK geometry 
\be
ds^2_\mathrm{KK} = -dt^2 + ds^2_\mathrm{TN}\,,
\label{qone}
\ee
where $ds^2_\mathrm{TN}$ is  the Taub-NUT metric
\be
ds^2_\mathrm{TN}= H^{-1} (dy - \tilde Q \cos\theta d\phi)^2 + H (d\rho^2 + \rho^2 d\theta^2 + \rho^2 \sin^2\theta d\phi^2)\,,\quad H = 1+{\tilde Q\over \rho}\,.
\label{taubnut}
\ee
Note that the time direction is trivially tensored with the Taub-Nut part of the metric. (The other five spatial directions of string theory are also assumed to be flat and directly tensored with this solution; we do not write them since they are not involved in the process of interest.)

Suppose we have a string wrapping the $y$ circle. A string winding around this circle creates a 2-form field $B^{(2)}$; the field strength of this gauge field gives a charge which can be measured from infinity. But the $y$ circle shrinks to zero at the core of the KK monopole. So we can unwind the string by moving it to the center of the monopole. What happens to the charge describing the winding number of the string? In \cite{ghm} it was shown that the process of unwinding the string creates a 2-form gauge field
\be
B^{(2)}_\mathrm{KK}= t \,d (H^{-1}  (dy  - \tilde Q \cos\theta d\phi))\equiv t\,\Omega^{(2)}_\mathrm{TN}\,.
\ee
This gauge field corresponds to a  field strength
\be
H^{(3)}_\mathrm{KK}= dt \wedge \Omega^{(2)}_\mathrm{TN} = dt\wedge [-H^{-2} dH \wedge (dy- \tilde Q \cos\theta d\phi) + H^{-1} \tilde Q \sin\theta d\theta\wedge d\phi]\,.
\label{tnh3}
\ee
This field strength gives the same charge at infinity which one would have obtained from the original $H^{(3)}$ created by the winding string. Thus charge is indeed conserved in the `unwinding' process. But to be able to carry the needed charge after the string has disappeared the geometry must have a topology which can support an appropriate $H^{(3)}$ field without source.  For the metric (\ref{taubnut}) the topologically non-trivial structure leads to the existence of an anti-self-dual closed 2-form
$\Omega^{(2)}_\mathrm{TN}=- *_4 \Omega^{(2)}_\mathrm{TN}$, and we have $B^{(2)}=t \Omega^{(2)}_\mathrm{TN}$. Thus
\be
*_5 H^{(3)}_\mathrm{KK} = *_5 (dt\wedge \Omega^{(2)}_\mathrm{TN})=- *_4 \Omega^{(2)}_\mathrm{TN}=\Omega^{(2)}_\mathrm{TN}
\label{qtwo}
\ee
and the equation of motion $d*_5 H^{(3)}_\mathrm{KK} =0$ is seen to be satisfied.

\subsection{Unwinding strings in D1D5 microstates}\label{method}

As discussed above, the D1D5 geometries have a curve ${\cal S}$ along which each point is the center of a KK monopole. Thus a string wound around the $S^1$ of a microstate solution can unwind if it touches any point on ${\cal S}$. This process will have to create a field $H^{(3)}$ on the microstate solution which will exhibit the charge originally carried by the wound string. But we cannot directly use the solution (\ref{tnh3}) for our problem. For one thing, we need the field $H^{(3)}$ in the geometry (\ref{three}), which is different (and more complicated) than (\ref{qone}). But even near the curve ${\cal S}$ where the topology is similar to that of a KK monopole,  our solution for $H^{(3)}$ will not reduce to the solution of \cite{ghm}. This is because the metric is not in general exactly that of a KK monopole, and more importantly, because 
the metric for the time direction (\ref{qthree}) is not  a simple tensor product with the spatial directions.  In particular the redshift $g_{tt}$ goes to zero at the curve ${\cal S}$. 

Thus we will have to find a different set of tools to generate the required solutions. We will see that supergravity fields other than $H^{(3)}$ need to be excited as well. But the underlying physics remains the same as that in \cite{ghm}: unwinding of a string results in a gauge field living on the microstate solution. 

\subsubsection{The nature of the supergravity perturbation}\label{nature}

The D1D5 microstate solutions actually allow a {\it set} of strings that can unwind. We can wrap an NS1 around the $S^1$ in (\ref{qfour}), but we could also have started with an NS5 and wrapped it on $T^4\times S^1$. This NS5 would give an effective string in the 6-d spacetime obtained by compactification on $T^4$. More generally, we could start with a bound state of $n_\mathrm{NS1}$ NS1 branes and $n_\mathrm{NS5}$ NS5 branes; this would also give an effective string in 6-d which we can unwind. We will always regard the unwinding string as a test string on our microstate background, and so write the resulting fields as linear perturbations to the microstate solution. Thus the fields resulting from the above bound state are $n_\mathrm{NS1}$ times the fields created by unwinding the $NS1$ plus $n_\mathrm{NS5}$ times the fields obtained by unwinding the NS5. Thus we need only find the fields generated by the unwinding of a single NS1 or a single NS5.

What happens when a string unwinds? The dynamical process of unwinding creates gauge fields, which remain nonzero after the string has disappeared. 
The actual solution after the unwinding process depends on the details of the unwinding process. We can have the following effects:

\b

(a) There will need to be a nonzero $H^{(3)}$ field carrying the required gauge charge. If we choose a solution carrying this gauge charge with the least possible energy, then we will have a BPS solution of the supergravity equations. This BPS solution will in general require other supergravity fields to be excited as well.

\b

(b) The perturbation can shift the 2-charge D1D5 solution to a neighboring D1D5 solution. The set of all D1D5 microstates is given by considering all vibration profiles for the NS1 string when we dualize D1D5 to NS1P. The vibrations of this NS1 can be in the noncompact directions (which generates the solutions (\ref{three})) or in the compact $T^4$ directions, which creates solutions studied in \cite{lmm,kst}. 

\b

(c) The process of unwinding can create additional energy above the minimum required to carry the winding charge; this extra energy will eventually flow out of the microstate to infinity.

\b

In \cite{ghm} a dynamical path from an initial (wound string) configuration to a final (unwound string) configuration was studied. But as can be seen from the above list, there are in general infinitely many different dynamical evolutions that lead to the unwinding of the string. Thus we will aim to just construct the final state solutions with minimum energy and the required charge; any other solution will have extra energy over these solutions, and this extra energy will leak off to infinity in due course. As we see from (b) in the above list, the BPS solution we write will also not be unique, since to a solution carrying the given string charge we can add  any  perturbation that shifts the initial D1D5 solution to a neighboring solution. 

\subsubsection{Methods for finding the solutions}

Thus our goal is to find a solution of supergravity that takes one of the nonsingular supergravity solutions (\ref{three}) and adds an infinitesimal perturbation carrying the charge of a NS1 wound along the $S^1$ direction. 
The essence of the method is the `boost' transformation, described as follows. Suppose we have a brane perpendicular to a direction $y$. If we boost in the direction $y$, then this brane picks up an additional charge -- the momentum charge in the direction $y$. Dualities can convert this momentum charge to a different kind of charge. The same process performed on the supergravity solution for the brane  gives the supergravity description of the brane plus momentum charge. Since our goal is to add an infinitesimal amount of NS1 charge, we will be performing an infinitesimal boost. We will also perform dualities before and after the boost to get the charge that we need. Different chains of these dualities will add  NS1 or NS5 charge to the D1D5 microstate solution. 

We will perform the required dualities and boosts in the next section. After that we examine another method of adding infinitesimal NS1 and NS5 charges. Consider the starting D1D5 solution, and perform an infinitesimal S-duality. This S-duality rotates the initial D1 charge to D1 plus an infinitesimal amount of NS1, and also rotates the initial D5 charge to D5 plus an infinitesimal amount of NS5. So by this method we cannot get any arbitrary ratio of NS1 and NS5 charges: these charges will appear in the same proportion that we had D1 and D5 charges in the starting solution. Thus while this method is much simpler to use, it generates only a 1-parameter subfamily of solutions out of the full 2-paramter family obtained by the first method. We verify that the solution obtained by this second method indeed agrees with the correct linear combination of solutions generated by the first method, up to the addition of an infinitesimal shift of D1D5 solution of type (b).

\section{Generating charge through boosts and dualities}
\setcounter{equation}{0}

We first add infinitesimal NS1 charge to the microstate solution (\ref{three}). Next, we take the result and perform dualities to map the added NS1 charge to an NS5 charge. The general solution obtained from unwinding a bound state of  NS1 and NS5 strings is given by superposing the two results.

\subsection{Adding infinitesimal NS1 charge}

Let us start from the solution (\ref{three}). Consider the following chain of dualities and boosts:
\be
\pmatrix{D1_y \cr D5_{y1234}}\stackrel{T_y}{\longrightarrow}\pmatrix{D0 \cr D4_{1234}}\stackrel{\mathrm{Boost}_y}{\longrightarrow}\pmatrix{D0 \cr D4_{1234}\cr P_y}\stackrel{T_y}{\longrightarrow}\pmatrix{D1_y \cr D5_{y1234}\cr N\!S1_y}\,.
\label{chain}
\ee

We list the rules for T and S duality in Appendix \ref{aone}. 
To perform the T-dualities we will 
adopt the `democratic' formalism: the field strengths have all allowed ranks up to the dimension of spacetime. Thus for the solution (\ref{three}) we need to construct  the 7-form dual to $F^{(3)}$:
\be
F^{(7)}=-* F^{(3)}\,.
\ee
One finds
\be
F^{(7)}=[d[H_5^{-1} (dt-A)\wedge (dy+B)] - *_4 dH_1] \wedge dz^4\,,
\ee
with $dz^4=dz_1\wedge\ldots \wedge dz_4$.

Let us now list the solution obtained after each step of the chain of dualities (\ref{chain}).

\subsubsection{T-duality along $y$}
After the first T-duality one finds
\bea
&&ds^2 = -(H_1 H_5)^{-1/2}(dt-A)^2 +(H_1 H_5)^{1/2} [dy^2+ dx_i dx_i] + \Bigl({H_1\over H_5}\Bigr)^{1/2} dz_a dz_a\,,\\
&&B^{(2)}=B\wedge dy\,,\quad e^{2\Phi}={H_1\over H_5}(H_1 H_5)^{1/2}\,,\nn
&&F^{(2)}=dH_1^{-1}\wedge (dt-A)-H_1^{-1} dA\,,\quad F^{(4)}=-H_1^{-1} dB \wedge (dt-A)\wedge dy- *_4 dH_5\wedge dy\,,\nn
&&F^{(6)}=[dH_5^{-1}\wedge (dt-A)-H_5^{-1} dA]\wedge dz^4\,,\nn
&&F^{(8)}=[-H_5^{-1} dB \wedge (dt-A)\wedge dy- *_4 dH_1\wedge dy]\wedge dz^4\nonumber \,.
\eea

\subsubsection{Boost along $y$}

We perform an infinitesimal boost along the direction $y$, with infinitesimal parameter $\theta$:
\be
t\to t+\theta y\,,\quad y\to y + \theta t\,,
\ee
and  keep only terms of first order in $\theta$. As described in Appendix \ref{aone}, when we perform  a T-duality it is convenient to write the metric using the vielbien in the direction of the duality coordinate. 
After the above boost the vielbein corresponding to the coordinate $y$ is 
\be
d\hat{y}=dy+\theta dt - \theta (H_1 H_5)^{-1}(dt-A)\,.
\ee

The solution after the boost is
\bea
&&ds^2= -(H_1 H_5)^{-1/2} (dt-A)^2 + (H_1 H_5)^{1/2} [d\hat{y}^2 + dx_i dx_i]+ \Bigl({H_1\over H_5}\Bigr)^{1/2} dz_a dz_a\,,\\
&&B^{(2)}=B\wedge d\hat{y} + \theta (H_1 H_5)^{-1} B \wedge (dt-A)\,,\quad e^{2\Phi}={H_1\over H_5}(H_1 H_5)^{1/2}\,,\nn
&&F^{(2)}=dH_1^{-1}\wedge (dt-A)-H_1^{-1} dA+\theta dH_1^{-1} \wedge d\hat{y}\,,\nn
&&F^{(4)}=-H_1^{-1} dB \wedge (dt-A)\wedge d\hat{y}- *_4 dH_5\wedge d\hat{y} - \theta (H_1 H_5)^{-1} *_4 dH_5 \wedge (dt-A)\,,\nn
&&F^{(6)}=[dH_5^{-1}\wedge (dt-A)-H_5^{-1} dA+\theta dH_5^{-1} \wedge d\hat{y}]\wedge dz^4\,,\nn
&&F^{(8)}=[-H_5^{-1} dB \wedge (dt-A)\wedge d\hat{y}- *_4 dH_1\wedge d\hat{y} - \theta (H_1 H_5)^{-1} *_4 dH_1 \wedge (dt-A)]\wedge dz^4\,.\nonumber
\eea

\subsubsection{T-duality along $y$}

The last T-duality gives (we list only the physical field strengths with rank $p\le5$)
\bea
&&ds^2 = (H_1 H_5)^{-1/2}[-(dt-A)^2 + (dy+B)^2]+(H_1 H_5)^{1/2} dx_i dx_i + \Bigl({H_1\over H_5}\Bigr)^{1/2} dz_a dz_a\,,\nn
&&B^{(2)}= \theta [dt\wedge dy - (H_1 H_5)^{-1} (dt-A)\wedge (dy+B)]\,,\quad e^{2\Phi}={H_1\over H_5}\,,\nn
&&F^{(1)}= \theta dH_1^{-1}\,,\quad F^{(3)}=d[H_1^{-1}(dt-A)\wedge (dy+B)]- *_4 dH_5\,,\nn
&&F^{(5)}= \theta [-(H_1 H_5)^{-1} *_4 dH_5 \wedge (dt-A)\wedge (dy+B)+ dH_5^{-1} \wedge dz^4]\,.
\label{ns1}
\eea

This is the solution that we wished to obtain. It describes an infinitesimal NS1 charge added to the D1D5 system when an NS1 string unwinds in the fuzzball geometry. It can be checked that the above solution satisfies the supergravity equations of motion at first order in $\theta$. 

\subsubsection{Measuring the NS1 charge}\label{measuring}

We have from the equations of motion
\be
d\Bigl[e^{-2\Phi} * H^{(3)} + C^{(0)}\wedge * F^{(3)}+C^{(2)}\wedge F^{(5)}-{1\over 2} C^{(2)}\wedge C^{(2)}\wedge H^{(3)}\Bigr]=0\,.
\label{qsix}
\ee
The NS1 charge of the solution is obtained by integrating the closed 7-form that appears in square brackets in the above equation over $S^3\times T^4$, where $S^3$ is the sphere at spatial infinity. The NS5 charge is given by the integral of $H^{(3)}$ over $S^3$.

When, as in our case, the RR gauge fields vanish at infinity, only the term proportional to $*H^{(3)}$ in eq. (\ref{qsix}) contributes to the charge. Hence, the NS1 and NS5 charges of the solution (\ref{ns1}) are 
\be
Q^{(1)}_\mathrm{NS1} ={1\over (2\pi)^2 V_4}\int_{S^3\times T^4}e^{-2\phi}*H^{(3)}=  \theta (Q_1 + Q_5) \,,
\label{chargens1}
\ee
\be
Q^{(1)}_\mathrm{NS5} = {1\over (2\pi)^2} \int_{S^3} H^{(3)}=0\,.
\ee
The number $n_\mathrm{NS1}$ of NS1 branes is given by $Q_\mathrm{NS1}$ through
\be
Q_\mathrm{NS1} =  {(2\pi)^4 g^2 \alpha'^3\over V_4} n_\mathrm{NS1}\,.
\ee

\subsection{Adding infinitesimal NS5 charge}

In the 6-d theory compactified on $T^4$ there is another kind string, which carries magnetic charge with respect to the NS 2-form field: this is, in the full 10D theory, an NS5 brane wrapping $S^1\times T^4$.  A solution carrying this type of charge can be obtained from the solution (\ref{ns1}) via the following sequence of dualities
\be
\pmatrix{D1_y \cr D5_{y1234}\cr N\!S1_y}\stackrel{S}{\longrightarrow}\pmatrix{N\!S1_y \cr N\!S5_{y1234}\cr D1_y}\stackrel{T_{1234}}{\longrightarrow}\pmatrix{N\!S1_y \cr N\!S5_{y1234}\cr D5_{y1234}}\stackrel{S}{\longrightarrow}\pmatrix{D1_y \cr D5_{y1234}\cr N\!S5_{y1234}}\,.
\label{chain2}
\ee
Under these dualities the D1D5 solution returns to itself, while the NS1 charge changes to an NS5 charge.

To perform the first S-duality one needs the RR gauge fields $C^{(0)}$ and $C^{(2)}$ for the solution (\ref{ns1}). The field strengths imply the potentials
\be
C^{(0)}=\theta (H_1^{-1}-\alpha)\,,\quad C^{(2)}=-dt\wedge dy+H_1^{-1} (dt-A)\wedge (dy+B)+C_5\,,
\ee
where $C_5$ is the 2-form dual to $H_5$ 
\be
d C_5 = - *_4 d H_5\,.
\ee
The arbitrary constant $\alpha$ reflects the freedom to add a constant to $C^{(0)}$ at infinity. Our unperturbed solution has $C^{(0)}=0$ at infinity. Since the action of string unwinding cannot change $C^{(0)}$ at infinity, we must choose $\alpha$ so that $C^{(0)}$ continues to vanish at infinity after the perturbation. This sets $\alpha=1$. 

We now list the solution obtained after each step of the chain of dualities (\ref{chain2}).

\subsubsection{S-duality}

\bea
&&ds^2 = H_1^{-1}[-(dt-A)^2 + (dy+B)^2]+H_5 \,dx_i dx_i +  dz_a dz_a\,,\nn
&&B^{(2)}=-dt\wedge dy+ H_1^{-1} (dt-A)\wedge (dy+B)+C_5
\,,\quad e^{2\Phi}={H_5\over H_1}\,,\nn
&&C^{(0)}= -\theta H_5^{-1}(1- H_1)\,,\quad C^{(2)}=\theta [-dt\wedge dy + (H_1 H_5)^{-1} (dt-A)\wedge (dy+B)] \,,\nn
&&F^{(5)}= \theta [-(H_1 H_5)^{-1} *_4 dH_5 \wedge (dt-A)\wedge (dy+B)+ dH_5^{-1} \wedge dz^4]\,.
\eea
The values of the gauge field strengths, in the democratic formalism, needed for the subsequent T-duality are:
\bea
&&F^{(1)}= -\theta [dH_5^{-1}- d(H_1 H_5^{-1})] \,,\\
&&F^{(3)}= \theta [H_1^{-1} dH_5^{-1}\wedge (dt-A)\wedge (dy+B)+H_5^{-1}  *_4 d H_5\nn
&&\quad\quad\quad +  H_1 H_5^{-1} (d(H_1^{-1} (dt-A)\wedge (dy+B)) - *_5 dH_5)]\,,\nn
&&F^{(7)}=\theta[-H_1^{-1} dH_5^{-1}\wedge (dt-A)\wedge (dy+B) -H_5^{-1} *_4 d H_5\nn
&&\quad\quad\quad +(d(H_5^{-1} (dt-A)\wedge (dy+B)) -*_4 dH_1)]\wedge dz^4\,,\nn
&&F^{(9)}=\theta [(H_1 H_5)^{-1} *_4 d H_5+ (H_1^{-1} *_4 dH_1 - H_5^{-1} *_4 dH_5)] \wedge (dt-A)\wedge (dy+B)\wedge dz^4\,.\nonumber
\eea

\subsubsection{T-duality along $z_1, z_2, z_3, z_4$}
\bea
&&ds^2 = H_1^{-1}[-(dt-A)^2 + (dy+B)^2]+H_5 \,dx_i dx_i +  dz_a dz_a\,,\nn
&&B^{(2)}=-dt\wedge dy+H_1^{-1} (dt-A)\wedge (dy+B)+C_5
\,,\quad e^{2\Phi}={H_5\over H_1}\,,\quad F^{(1)}=\theta  dH_5^{-1}\,,\nn
&&F^{(3)}=\theta[-H_1^{-1} dH_5^{-1}\wedge (dt-A)\wedge (dy+B) -H_5^{-1} *_4 d H_5\nn
&&\quad\quad\quad + (d(H_5^{-1} (dt-A)\wedge (dy+B))-*_4 dH_1)]\,,\nn
&&F^{(5)}=-\theta [dH_5^{-1}- d (H_1 H_5^{-1})] \wedge dz^4\nn
&&\quad\quad\quad +\theta[(H_1 H_5)^{-1} *_4 d H_5+ (H_1^{-1} *_4 dH_1 - H_5^{-1} *_4 dH_5)]\wedge (dt-A)\wedge (dy+B)\,.\nn
\eea

The gauge fields $C^{(0)}$ and $C^{(2)}$ corresponding to $F^{(1)}$ and $F^{(3)}$ are
\bea
&&C^{(0)}=\theta (H_5^{-1}-\beta)\,,\\
&&C^{(2)}=\theta [-\beta dt\wedge dy+(\beta H_1^{-1}+ H_5^{-1} -(H_1 H_5)^{-1}) (dt-A)\wedge (dy+B)+ C_1 +\beta C_5]\,,\nonumber
\eea
where
\be
d C_1 = -*_4 d H_1 \,,\quad d C_5 = -*_4 d H_5\,.
\ee
Here the parameter $\beta$ has appeared in a manner similar to $\alpha$. Requiring the vanishing of $C^{(0)}$ at infinity, we find $\beta=1$.

\subsubsection{S-duality}
\bea
&&\!\!\!\!ds^2 = (H_1 H_5)^{-1/2}[-(dt-A)^2 + (dy+B)^2]+(H_1 H_5)^{1/2} dx_i dx_i + \Bigl({H_1\over H_5}\Bigr)^{1/2} dz_a dz_a\,,\nn
&&\!\!\!\!B^{(2)}=\theta [- dt\wedge dy+(H_1^{-1}+ H_5^{-1} -(H_1 H_5)^{-1}) (dt-A)\wedge (dy+B)+ C_1 +C_5]\,,\,\,\,e^{2\Phi}={H_1\over H_5}\,,\nn\nn
&&\!\!\!\!F^{(1)}=-\theta d [H_1^{-1}(1-H_5)]\,,\quad F^{(3)}=-d[H_1^{-1}(dt-A)\wedge (dy+B)]+ *_4 dH_5\,,\nn
&&\!\!\!\!F^{(5)}=-\theta [dH_5^{-1}- d (H_1 H_5^{-1})] \wedge dz^4\nn
&&\quad\quad\quad +\theta[(H_1 H_5)^{-1} *_4 d H_5+(H_1^{-1} *_4 dH_1 - H_5^{-1} *_4 dH_5)]\wedge (dt-A)\wedge (dy+B)\,.
\eea

In the solution above the overall sign of the field $F^{(3)}$ is the opposite of the one in solution (\ref{ns1}) (this happens because the product of two S-dualities reverses the sign of $B^{(2)}$ and $C^{(2)}$). To compare with the solution (\ref{ns1}), it is convenient to reverse the sign of both $F^{(3)}$ and $H^{(3)}$, leaving the signs of the other fields invariant (it is easy to check that this is a symmetry of the theory). Hence the final solution is
\bea
&&\!\!\!\!ds^2 = (H_1 H_5)^{-1/2}[-(dt-A)^2 + (dy+B)^2]+(H_1 H_5)^{1/2} dx_i dx_i + \Bigl({H_1\over H_5}\Bigr)^{1/2} dz_a dz_a\,,\nn
&&\!\!\!\!B^{(2)}=\theta [dt\wedge dy+((H_1 H_5)^{-1}-H_1^{-1}- H_5^{-1}) (dt-A)\wedge (dy+B)-C_1 - C_5]\,,\,\,\,e^{2\Phi}={H_1\over H_5}\,,\nn
&&\!\!\!\!F^{(1)}=-\theta d [H_1^{-1} (1- H_5)]\,,\quad F^{(3)}=d[H_1^{-1}(dt-A)\wedge (dy+B)]- *_4 dH_5\,,\nn
&&\!\!\!\!F^{(5)}=-\theta [dH_5^{-1}- d (H_1 H_5^{-1})] \wedge dz^4\nn
&&\quad\quad\quad +\theta[(H_1 H_5)^{-1} *_4 d H_5+ (H_1^{-1} *_4 dH_1 - H_5^{-1} *_4 dH_5)]\wedge (dt-A)\wedge (dy+B)\,.
\label{ns5}
\eea
This is the solution that we wished to obtain for the unwinding of an NS5 brane in the fuzzball geometry.

\subsubsection{Measuring the NS5 charge}

Following the same steps as we did in  section \ref{measuring}, we find that the solution
(\ref{ns5}) carries the charges 
\be
Q^{(5)}_\mathrm{NS1} = {1\over (2\pi)^2 V_4} \int_{S^3\times T^4} *H^{(3)}= 0\,,
\ee
\be
 Q^{(5)}_\mathrm{NS5} = {1\over (2\pi)^2} \int_{S^3} H^{(3)}=\theta (Q_1 + Q_5) \,.
\label{chargens5}
\ee
The number $n_\mathrm{NS5}$ of NS5 branes is given by $Q_\mathrm{NS5}$ through
\be
Q_\mathrm{NS5} = \alpha' n_\mathrm{NS5}\,.
\ee

\section{Adding NS1-NS5 charges by infinitesimal S-duality}
\setcounter{equation}{0}

In this section we will look at a second method for generating the fields resulting from the unwinding of a bound state of NS1 and NS5. As mentioned in section \ref{method}, this method will only give solutions that have the NS1 and NS5 charges in a certain proportion. 

We will take the D1D5 solution (\ref{three}) and apply an infinitesimal S-duality. This generates NS1-NS5 charges in proportion to the D1-D5 charges already present in the solution. 

Let us now carry out the required steps. Infinitesimal S-duality does not change the metric at linear order. It also leaves the 5-form $F^{(5)}$ invariant. It acts on the complex scalar
$\tau = C^{(0)} + i e^{-\Phi}$ as
\be
\tau \to {\tau -\theta\over \theta \tau +1}\,.
\ee
The starting solution (\ref{three}) has $\tau = i e^{-\Phi}$ and thus the solution after S-duality has
\be
\tilde \Phi= \Phi \,,\quad \tilde C^{(0)} = \theta (e^{-2\Phi}-1)= \theta \Bigl({H_5\over H_1}-1\Bigr)\,.
\ee

The NSNS and RR 2-forms transform as
\be
\pmatrix{B^{(2)}\cr C^{(2)}}\to \pmatrix{\tilde B^{(2)}\cr \tilde C^{(2)}} = \pmatrix{1& -\theta\cr \theta & 1} \pmatrix{B^{(2)}\cr C^{(2)}}\,,
\ee
which implies
\be
\tilde B^{(2)} = -\theta C^{(2)}\,,\quad \tilde C^{(2)}= C^{(2)}\,.
\ee

In what follows we will drop all terms of order $\theta^2$. We find that the solution generated by infinitesimal S-duality is
\bea
ds^2 &=& (H_1 H_5)^{-1/2}[-(dt-A)^2 + (dy+B)^2]+(H_1 H_5)^{1/2} dx_i dx_i + \Bigl({H_1\over H_5}\Bigr)^{1/2} dz_a dz_a\,,\nonumber\\
F^{(3)}&=&d[H_1^{-1} (dt -A)\wedge (dy +B)]- *_4 dH_5\,,\quad e^{2\Phi}={H_1\over H_5}\,,\nonumber\\
B^{(2)}&=& -\theta(H_1^{-1} (dt -A)\wedge (dy +B) + C_5)\,,\quad F^{(1)} = \theta d  \Bigl({H_5\over H_1}\Bigr)\,,\nonumber\\
F^{(5)}&=&0\,.
\label{chargesduality}
\eea

The charges carried by this solution are
\be
Q_\mathrm{NS1} ={1\over (2\pi)^2 V_4}\int_{S^3\times T^4}e^{-2\phi}*H^{(3)}=  \theta Q_1 \,,
\label{qqone}
\ee
\be
Q_\mathrm{NS5} = {1\over (2\pi)^2} \int_{S^3} H^{(3)}=\theta Q_5 \,.
\label{qqtwo}
\ee

\subsection{Generating the same charges by the first method}

We can generate a solution with charges (\ref{qqone}),(\ref{qqtwo}) by adding together solutions of type (\ref{ns1}) and type (\ref{ns5}). Looking at the charges (\ref{chargens1}),(\ref{chargens5}) we see that we should add
the solutions (\ref{ns1}) and (\ref{ns5}) with coefficients proportional to $Q_1$ and $Q_5$. Let us call the result the `solution obtained by the first method'.

This solution obtained by the first method will now carry the same charges as the solution (\ref{chargesduality}). Let us subtract (\ref{chargesduality}) from the solution obtained by the first method. This difference will carry  no charges, and is found to be nonzero
 \bea
ds^2 &=& (H_1 H_5)^{-1/2}[-(dt-A)^2 + (dy+B)^2]+(H_1 H_5)^{1/2} dx_i dx_i + \Bigl({H_1\over H_5}\Bigr)^{1/2} dz_a dz_a\,,\nonumber\\
F^{(3)}&=&d[H_1^{-1} d\hat t \wedge d\hat y]- *_4 dH_5\,,\quad e^{2\Phi}={H_1\over H_5}\,,\nonumber\\
B^{(2)}&=& \theta\Bigl\{ [(Q_5-Q_1) (H_1 H_5)^{-1} + Q_1 H_1^{-1} - Q_5 H_5^{-1}]\, d\hat t \wedge d\hat y + Q_1 C_5 - Q_5 C_1\Bigr\}\,,\nonumber\\
F^{(1)}&=& \theta\Bigl\{ (Q_1-Q_5) dH_1^{-1}-Q_1\, d\Bigl({H_5\over H_1}\Bigr)\Bigr\}\,,\nonumber\\
F^{(5)}\!\!\!&=&\!\!\! \theta\Bigl\{ [(Q_5-Q_1) (H_1 H_5)^{-1} *_4 d H_5 + Q_5 (H_1^{-1}*_4 d H_1 - H_5^{-1} *_4 d H_5)]\,\wedge d\hat t \wedge d\hat y\nonumber\\
&& - \Bigl[(Q_5-Q_1)d H_5^{-1} - Q_5 \,d\Bigl({H_1\over H_5}\Bigr)\Bigr] \wedge dz^4\Bigr\}\,,
\label{hair}
\eea
where we have written
\be
d\hat t = dt -A \,,\quad d\hat y = dy +B\,.
\ee

How should we interpret this nonzero difference between two solutions carrying the same charges? Recall that we had noted in section \ref{method} that the fields generated by string unwinding are unique only up to the addition of further perturbations that can change the starting D1D5 solution to a neighboring solution (type (b) in the list of perturbations). Thus we should ask if (\ref{hair}) is a perturbation that changed the D1D5 solution (\ref{three}) to another extremal, infinitesimally different D1D5 solution. We will now show that this is indeed the case.

\subsection{Analyzing the perturbation (\ref{hair})}

As mentioned in section \ref{solution}, the extremal D1D5 solutions can be obtained by S,T dualities applied to the states of the NS1P system, which is just a string carrying a vibration profile. These vibrations can be in the four noncompact directions; this vibration profile is given by $\vec F(v)$ and leads to the solutions (\ref{three}). But we can also have vibrations in the $T^4$ directions. The latter vibrations are given by profiles ${\mathcal{F}}(v), {\mathcal{F}}_\alpha(v)$, where we break up the four torus directions into a special one and three others (labelled by $\alpha$). The reason for this breakup is that one of the four torus directions is used to perform a T-duality when mapping from NS1P to D1D5, and so the profile in this direction appears differently in the final D1D5 solution.

We let ${\mathcal{F}}(v)$ be small;  we will work to linear order in this variable. We set ${\mathcal{F}}_\alpha(v)=0$.  Adding such a vibration profile to the vibrations already encoded in $\vec F$, we get a D1D5 solution given by (\ref{three}) together with the additional fields \cite{kst}
\bea
B^{(2)}&=& {\mathcal{A}\over H_1 H_5} d\hat t \wedge d\hat y - \mathcal{B}\,,\label{skenderis1}\\
F^{(1)}&=& - d\Bigl[{\mathcal{A}\over H_1}\Bigr]\,,\label{skenderis2}\\
F^{(5)}&=&H_1^{-1} \Bigl({\mathcal{A}\over H_5}*_4 dH_5 - d\mathcal{B} \Bigr)\wedge d\hat t \wedge d\hat y+H_5^{-1} \Bigl({\mathcal{A}\over H_5} dH_5 - d\mathcal{A}\Bigr)\wedge dz^4\,.
\label{skenderis3}
\eea
Here $\mathcal{A}$, $\mathcal{B}$ are defined in a manner similar to (\ref{harmonicfunctions})
\be
\mathcal{A} = - {Q_5\over L_T} \int_0^{L_T}dv {\dot{\mathcal{F}}(v)\over |x-F(v)|^2}\,,
~~~~
*_4 d \mathcal{B} =d \mathcal{A}\,.
\ee

Suppose we take
\be
\mathcal{A} =  \theta[Q_1 (H_5-1) - Q_5 (H_1-1)]\,,
\label{choice}
\ee
and the following choice for $\mathcal{B}$ (which satisfies $*_4 d \mathcal{B} =d \mathcal{A}$)
\be
\mathcal{B}=\theta[Q_5 C_1 - Q_1 C_5]\,.
\ee
Then we find that the D1D5 perturbation (\ref{skenderis1}-\ref{skenderis3}) gives just the difference (\ref{hair}) between the solutions we had obtained by our two different methods. The choice (\ref{choice}) for $\mathcal{A}$ corresponds to taking a torus vibration profile
\be
\dot{\mathcal{F}}(v)= \theta Q_5 \,\Bigl[\Bigl({1\over L_T}\int_0^{L_T} dv' |\dot{F}(v')|^2 \Bigr)- |\dot{F}(v)|^2\Bigr]\,.
\ee
Note that the profile $\mathcal{F}(v)$ defined above is indeed a closed profile $\mathcal{F}(v+L_T)=\mathcal{F}(v)$, as it should be if we have a true bound state of NS1P (the NS1 string should form one closed loop \cite{what}). 

To summarize, we see that the method of infinitesimal S-duality gives a comparatively simple way of generating the fluxes resulting from the unwinding of an effective string in a D1D5 microstate geometry. While this method is simple, it allows us to only consider effective NS1-NS5 string bound states that have the NS1 and NS5 charges in the same proportion as the D1 and D5 charges in the initial microstate. We see that the result of this computation agrees with the results we would obtain by the first method that we studied (where the NS1 and NS5 charges could be chosen independently), up to the addition of an infinitesimal shift in the extremal D1D5 microstate.

\section{Comparing the D1D5 solutions with the KK monopole}\label{moore}
\setcounter{equation}{0}

We had noted in section \ref{near} that at the curve ${\cal S}$ in the D1D5 microstate the geometry is regular, even though the functions (\ref{harmonicfunctions}) appear to imply a singularity. The apparent singularity arises because the points on ${\cal S}$ are `centers of KK monopoles'. In section \ref{unwinding1} we had recalled the computation of \cite{ghm} where the $H^{(3)}$ produced by an unwinding string in a KK monopole was computed. One might at first think that the supergravity perturbation in the D1D5 microstates should reduce to the $H^{(3)}$ in the KK monopole computation, but such is not the case; the time direction in the D1D5 microstates appears in the metric differently than in the pure KK monopole case (where it is trivially tensored with the rest of the geometry). In this section we examine this difference in more detail.

Consider a D1D5 microstate geometry, and recall the discussion of section \ref{near} where we considered this geometry in the vicinity of the curve ${\cal S}$. We now wish to examine the harmonic functions in the metric a little further away from ${\cal S}$, to see the geometrical structure of the metric in the neighborhood of ${\cal S}$. Let a part $v=(0, v_1)$ of the curve lie in the segment which we are approximating locally by a straight line, and $v=(v_1, L_T)$ describe the remainder of ${\cal S}$.

There is actually a good separation between the contributions of these two parts of the curve ${\cal S}$. To see this, we recall the discussion in \cite{phase} where the nature of the generic fuzzball was studied. 
There are three length scales of interest, and we note their scalings with powers of $N_1N_5$. These are:  (a) the radius of the $S^1$ which fibers nontrivially to generate the KK monopole; this length is $\sim 1$ (b) The typical separation between a strand of ${\cal S}$ and its neighboring strand; this is $(N_1N_5)^{1\over 6}$ (c) The diameter of the fuzzball; this is $(N_1N_5)^{1\over 2}$. What is relevant to us is the ratio ${\cal R}$ of (b) to (a); this is large (since $N_1N_5$ is large). Thus the KK monopole structure of a fiber of ${\cal S}$ `finishes off' at a distance $\sim 1$ from the axis of the fiber, much before we encounter any interference from other strands of ${\cal S}$. We are of course not considering generic fuzzballs, but we can look at fuzzballs where  $1\ll {\cal R}\ll (N_1N_5)^{1\over 6}$. This provides a large collection of fuzzballs which can be described by classical geometries (due to the second inequality), and are the ones of interest to us in our present study. 

Let us now see how to write down the contributions to the harmonic functions (\ref{harmonicfunctions}) from the `near part' ($v=(0, v_1)$) and `far part' (the remainder) of ${\cal S}$.  The near part gives the usual structure of an isolated KK monople; thus we will get for example $H_5\approx A+{B\over \rho}$. The constant $B$ can be fixed by going near to the center of the KK monopole; thus these singular parts of the harmonic functions can be found from (\ref{hfone}),(\ref{hftwo}). The contribution of the `far part' of the curve will be approximately a constant in the region where the KK monople structure is nontrivial. To find these constants we will have to integrate over all the `far part' of ${\cal S}$, but here we need only the overall structure of the result:
\be
H_1 \approx \alpha_1+{\t Q_1\over \rho}\,\quad H_5 \approx \alpha_2+{\t Q_5\over \rho}\,,\quad A_i\approx \alpha_{3,i} dx_i-{\sqrt{\t Q_1\t Q_5}\over\rho} dz\,,
\label{hfthree}
\ee
where
\be
\t Q_1 = {\pi Q_5 |\dot{F}(v_0)|\over L_T}\,,\quad \t Q_5 = {\pi Q_5\over  |\dot{F}(v_0)| L_T}\,.
\ee
As shown in section \ref{near}, this geometry is nonsingular at $\rho\r 0$. The topological structure around $\rho=0$ is that of a KK monopole times a line. To see this more explicitly, let us take the charges and the constants in (\ref{hfthree})
to be
\be
\t Q_1=\t Q_5 \equiv \t Q\,,\quad \alpha_1=\alpha_2, ~~~\alpha_{3,i}=0\,.
\ee

In this case the rescalings $\rho\to \rho/\alpha_1$, $z\to z/\alpha_1$, plus an overall rescaling of the metric by the factor $\alpha_1$, bring the metric near ${\cal S}$ to the form (ignoring the $T^4$)
\be
ds^2_{\mathrm{near}\scriptscriptstyle{-}\mathcal{S}} = -H^{-1} (dt+{\t Q\over \rho} dz)^2 + H dz^2 + ds^2_\mathrm{TN}\,,
\label{metricone}
\ee
where $H=1+{\t Q\over \rho}$ and $ds^2_\mathrm{TN}$ is  the Taub-NUT metric 
\be
ds^2_\mathrm{TN}= H^{-1} (dy -\t Q \cos\theta d\phi)^2 + H (d\rho^2 + \rho^2 d\theta^2 + \rho^2 \sin^2\theta d\phi^2)\,.
\ee
As reviewed in section \ref{unwinding1}, the metric considered in \cite{ghm} is that of a KK monople, which has a trivial time direction
\be
ds^2_\mathrm{KK} = -dt^2 + ds^2_\mathrm{TN}\,.
\label{metrictwo}
\ee
The process of unwinding an NS1 in this KK monopole geometry produces the field
\be
H^{(3)}_\mathrm{KK}= dt \wedge \Omega^{(2)}_\mathrm{TN} = dt\wedge [-H^{-2} dH \wedge (dy- \tilde Q \cos\theta d\phi) + H^{-1} \tilde Q \sin\theta d\theta\wedge d\phi]\,.
\label{kkkk}
\ee
By contrast, the $H^{(3)}$ field in the D1D5 microstate near ${\cal S}$ turns out to have the form
\be
H^{(3)}_{\mathrm{near}\scriptscriptstyle{-}\mathcal{S}}\approx 2 H^{-3} dH \wedge (dt-A)\wedge (dy+B)+H^{-2} dA\wedge(dy+B) +H^{-2} (dt-A)\wedge dB\,, 
\label{rrrr}
\ee
with
\be
H= 1+ {\t Q\over \rho}\,,\quad A = - {\t Q\over \rho} dz\,,\quad B = - \t Q \cos\theta d\phi\,.
\ee
Note that the power of the harmonic function $H$ is different in (\ref{kkkk}) and (\ref{rrrr}). This difference can be traced to the fact that the metric in the D1D5 case (\ref{metricone}) has a different $t$ part than the metric (\ref{metrictwo}) in the KK monople case. 
Further, in the D1D5 microstate the fields $F^{(1)}$ and $F^{(5)}$ need to be nonzero as well to solve the equations of motion. This fact can be traced to the presence of the gauge field $F^{(3)}$ in the unperturbed D1D5 solution.

Despite these technical differences, the physics of string unwinding is the same in the two cases, since it arises from the topological structure of the  
geometry.

\section{Discussion}
\setcounter{equation}{0}

The extremal D1D5 system gives the simplest example of a black hole. Consider the extremal bound states of $n_1$ D1 branes and $n_5$ D5 branes. Suppose we dimensionally reduce to the 5-d noncompact spacetime, assume a spherically symmetric ansatz and solve the leading order classical supergravity equations with these charges. We find a solution that has a singularity at $r=0$. But if we include 1-loop corrections arising from strings winding around the  compact $S^1$, then we find higher derivative terms in the effective action. For the case of K3 compactifications it was shown in \cite{dabholkar,higher,senrecent} that the resulting higher-derivative theory has a solution where we get a nonzero Bekenstein-Wald entropy; this entropy equals the entropy given by the microscopic count of states. The horizon here is `small' since it arises only after higher derivative corrections are included, but if we were to associate a geometry to the solution it would have the structure of the extremal Reissner-Nordstrom hole -- an infinite throat, a horizon, and a singularity inside the horizon. 

The actual solutions of string theory with the given D1, D5 charges do not have  a horizon; they are best described in 10-d, where they are found to be `fuzzballs'. We have taken simple microstates for this system -- those given by solutions that are close to classical -- and studied the unwinding of a string in the microstate geometry. The geometry has the topology of a KK monopole times a curve ${\cal S}$ in its `cap' region; when a string with winding hits any point on this curve it can unwind, leaving its winding charge in the form of a gauge field strength on the geometry. This process is similar to the unwinding of a string in a KK monopole studied in \cite{ghm}. We considered NS1 strings as well as effective strings obtained by wrapping an NS5 on the compact $T^4$; linearity at leading order implies that this gives the result for all bound states of NS1 and NS5 branes. The fields produced on the microstate due to the unwinding were computed by two different methods. It was noted that while the topology of the D1D5 microstate has a local KK monopole like structure, the metric is actually quite different, and so the fields produced differ in form from the KK monopole case.

\begin{figure}[ht]
\begin{center}
\includegraphics[width=12cm]{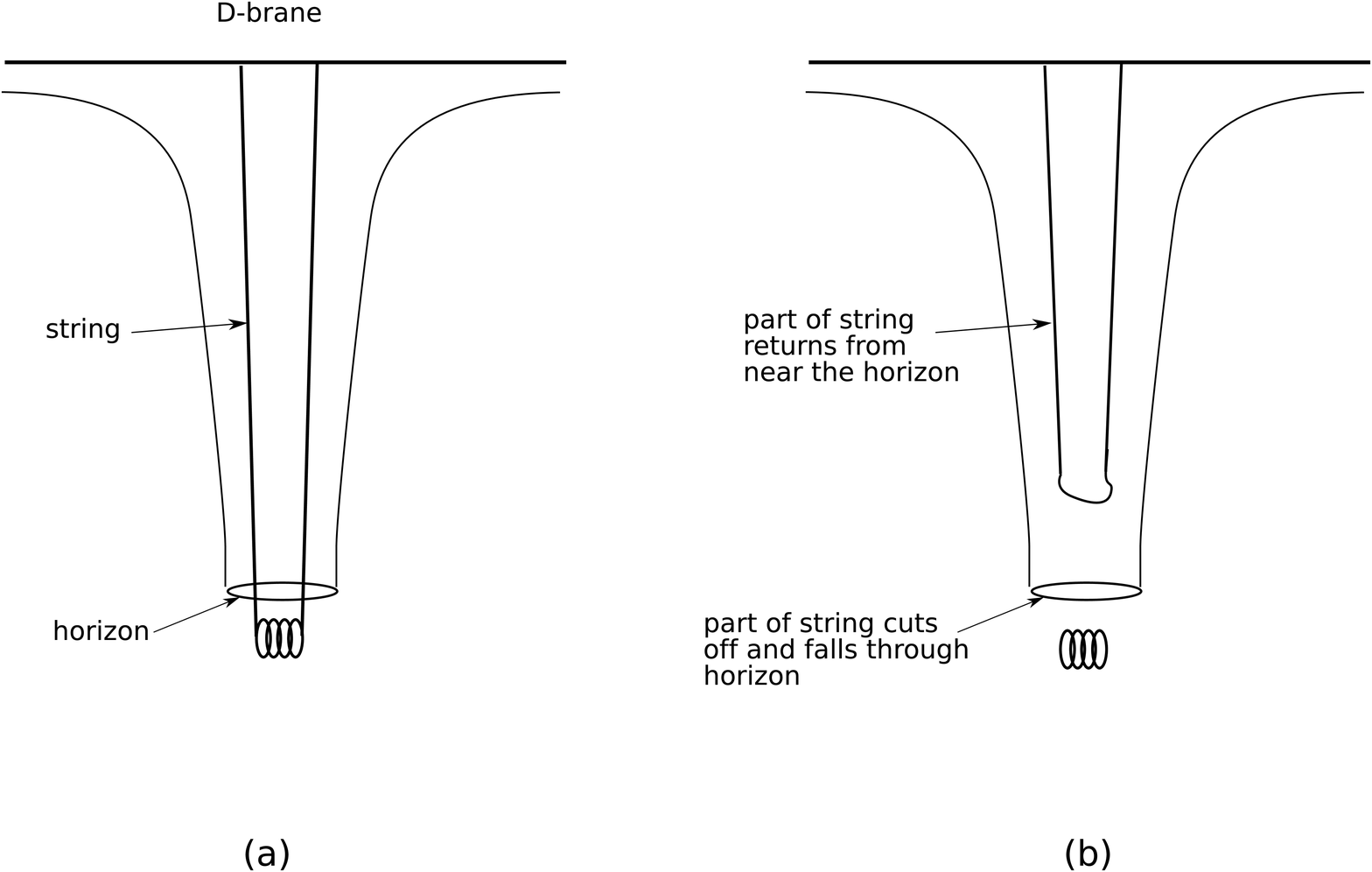}
\caption{(a) A string carrying winding number along $S^1$ falls in through the horizon of the naive black hole geometry; the hole will pick up the winding charge and the ends of the string will appear stuck on the horizon (b) The infalling string can self-iontersect and cut, with the part carrying winding charge falling through the horizon, and the rest retreating from the horizon with no winding.}\label{fatwo}
\end{center}
\end{figure}

Since the `naive geometry' of the D1D5 system gives a `small' black hole, it is not clear to what extent we should expect traditional black hole type behavior for this system. Here we will content ourselves with noting some relations between the actual evolution of a wound string in the fuzzball geometries and what one might see if the string fell into an extremal Reissner-Nordstrom geometry.  The naive geometry of an extremal hole has an infinite throat but an infalling object can cross the horizon in a finite proper time. A string carrying winding can cross the horizon; it would then appear that the winding charge has disappeared down the hole and the black hole has picked up the winding charge (fig.\ref{fatwo}(a)). Of course in this process an outside observer would still find that the string has two ends stuck on the horizon. (The `membrane paradigm' \cite{membrane} gives an effective boundary condition for flux lines when a charge falls through the horizon, and the gauge field created by the winding string should have a similar behavior.) One can also imagine a process where the infalling string self-intersects and cuts off, with a part carrying winding charge falling through the horizon (fig.\ref{fatwo}(b)). This process looks similar to the process in the microstate depicted in fig.\ref{faone}. It is not clear if we should compare these processes to evolution in the microstate geometries, but if we wanted to make such a comparison, we would have to first compute the probability of unwinding in a generic microstate. We have not done that in the present paper, restricting ourselves to showing that the string is allowed to unwind, and finding the field produced on the microstate when it does. 

Nevertheless, we can note  some similarities between the microstate computation and the expectation from infall in the naive geometry. In the naive geometry, the black hole `has no hair', and the charge swallowed by the hole gives a field that rapidly settles down (after decay of quasi-normal modes) to a spherical configuration. In our microstate, the field $H^{(3)}$ (and other fields as well) spread out all along the curve ${\cal S}$, which winds around in a complicated way inside the `cap' region of the geometry. This fact is not immediately obvious -- one might think that if an NS1 unwinds after touching a pint $P$ on the curve ${\cal S}$ then the $H^{(3)}$ field produced by the unwinding may be localized near $P$, and thus be located asymmetrically inside the `fuzzball'. If such were the case, we would have a large family of perturbations exhibiting the same NS1 charge, arising from different locations $P$. But we know that there cannot be such a family, since the extremal bound state of D1D5 plus an NS1 has the {\it same} entropy as just the D1D5. Thus there should be a unique solution for the supergravity perturbation resulting from the unwinding from the NS1 (up to shifts of the D1D5 extremal microstate and additions of nonextremal excitations, noted in section \ref{nature}). This is the solution that we have found.

But given that this excitation spreads along the curve ${\cal S}$, we find that for a generic microstate the field outside the fuzzball region will be spherically symmetric to a high order, since it will have very high multipole moments. (The curve ${\cal S}$ makes a random walk inside the fuzzball region \cite{phase}, with the length of the walk scaling like $(n_1n_5)^\h$.)
This argument is similar in spirit to the discussion in \cite{lm5} about why the boundary of the fuzzball acts like a horizon: if a quantum falls into the complicated fuzzball region, then because of the complicated nature of the geometry the ratio of the re-emergence time to the crossing time goes to infinity as $n_1n_5\r \mathrm{infinity}$. 
We should think of the large number $n_1n_5$ as a power of ${1\over \hbar}$ from a classical viewpoint, so  in a classical limit $\hbar\r 0$ the quantum would not emerge at all. Similarly, the multipole moment of the field strength in the present case would be infinity in a classical limit, so the field would look spherically symmetric for an outside observer, similar to what would be predicted by infall into the naive black hole geometry.

We now see a loose similarity between the evolution in the fuzzball microstate and evolution in the naive black hole geometry. The case of fig.\ref{fatwo}(a) where the string has ends stuck at the horizon corresponds to the situation where the string carrying winding has fallen into the fuzzball region and become trapped. The complexity of this region does not allow for a quick reemergence, similar to the case of quanta that fall into the complicated cap geometry \cite{lm5}. Thus even if the part of the string inside the fuzzball unwinds, the string may not be able to extricate itself from the cap region in a short time.  The case in fig. \ref{fatwo}(b) corresponds to the situation where the string has  unwound in the fuzzball and managed  to reemerge without winding; the winding charge is then carried by gauge fields on the fuzzball.\footnote{In \cite{ghm} it was noted that the string can also unwind while staying at a distance $r$ from the center of the KK monopole; this however requires the string to stretch to a length equal to the circumference of the sphere at radius $r$.  Thus in our fuzzball the string can unwind without actually hitting the center of the KK monopole tube, though the process may need a little more  energy since the string has to stretch more. {\it Outside} the fuzzball there is no KK monopole structure, since the KK monopole is a `dipole charge' in the geometry with total value zero; thus the string will need to enter the fuzzball region to be able to unwind by using the KK structure.} It should be noted that these are loose similarities between the fuzzball situation and the naive geometry, and more dynamical studies would be needed to check if the naive geometry has a meaningful role to play in providing an effective description of the generic fuzzball.\footnote{See for example \cite{bala,dasmandal} for studies relating the naive geometry to effective descriptions of fuzzballs.}

\section*{Acknowledgements}

We thank Steve Avery, Iosif Bena, Borun Chowdhury, Sumit Das, Sheer El Showk, Cl\'ement Ruef and Masaki Shigemori for interesting discussions. The work of SDM was supported in part by DOE grant DE-FG02-91ER-40690.

\appendix
\section{Conventions and duality rules}\label{aone}
\renewcommand{\theequation}{A.\arabic{equation}}
\setcounter{equation}{0}

In this Appendix we summarize the equations of supergravity and the rules for T and S dualities that we use.

\subsection{Gauge fields and field equations}

We work in type II string theory. The bosonic supergravity fields consist of   metric $g$,  the dilaton $\Phi$, the NSNS 2-form field, $B^{(2)}$  and the RR p-form fields $C^{(p)}$  ($p=1,3$ for IIA and $p=0,2,4$ for IIB). The  field strengths are defined by
\be
H^{(3)}=d B^{(2)}
\ee
and
\be
F^{(p+1)}= d C^{(p)}  \quad \mathrm{for}\,\,p< 2\,,\quad F^{(p+1)}= d C^{(p)}+ H^{(3)}\wedge C^{(p-2)} \quad \mathrm{for}\,\,p\ge 2\,.
\label{fieldstrengths}
\ee
The field strength definitions imply the Bianchi identities:
\bea
d H^{(3)}=0\,,\quad d F^{(p+1)}=0 \quad \mathrm{for}\,\,p< 2\,,\quad d F^{(p+1)} = - H^{(3)}\wedge F^{(p-1)}\quad \mathrm{for}\,\,p\ge 2\,.
\label{bianchi}
\eea
The equations of motion for type IIA are 
\bea
&&e^{-2\Phi} \Bigl(R_{mn}+2 \nabla_m \nabla_n \Phi-{1\over 4} H^{(3)}_{mpq}H^{(3)pq}_n\Bigr)-{1\over 2} F^{(2)}_{mp}F^{(2)p}_n-{1\over 2}{1\over 3!} F^{(4)}_{mpqr}F^{(4)pqr}_n\nn
&&\qquad+{1\over 4} g_{mn}\Bigl({1\over 2} F^{(2)}_{pq} F^{(2)pq} + {1\over 4!} F^{(4)}_{pqrs} F^{(4)pqrs} \Bigr)=0\,,\nn
&&4 d * d \Phi - 4 d\Phi \wedge * d\Phi + * R -{1\over 2} H^{(3)}\wedge * H^{(3)}=0\,,\nn
&&d * (e^{-2\Phi} H^{(3)})+ F^{(2)}\wedge * F^{(4)}+{1\over 2} F^{(4)}\wedge F^{(4)}=0\,,\nn
&& d * F^{(2)}+H^{(3)}\wedge * F^{(4)}=0\,,\quad d*F^{(4)}+ H^{(3)}\wedge F^{(4)}=0\,,
\label{eomA}
\eea
and for type IIB are
\bea
&&\!\!\!\!e^{-2\Phi} \Bigl(R_{mn}+2 \nabla_m \nabla_n \Phi-{1\over 4} H^{(3)}_{mpq}H^{(3)pq}_n\Bigr)-{1\over 2} F^{(1)}_{m}F^{(1)}_n-{1\over 2} {1\over 2!}F^{(3)}_{mpq}F^{(3)pq}_n-{1\over 4}{1\over  4!} F^{(5)}_{mpqrs}F^{(5)pqrs}_n\nn
&&\qquad+{1\over 4} g_{mn}\Bigl(F^{(1)}_{p} F^{(1)p} + {1\over 3!} F^{(3)}_{pqr} F^{(3)pqr} \Bigr)=0\,,\nn
&&\!\!\!\!4 d * d \Phi - 4 d\Phi \wedge * d\Phi + * R -{1\over 2} H^{(3)}\wedge * H^{(3)}=0\,,\nn
&&\!\!\!\!d * (e^{-2\Phi} H^{(3)})+ F^{(1)}\wedge * F^{(3)}+F^{(3)}\wedge F^{(5)}=0\,,\nn
&& \!\!\!\!d * F^{(1)}-H^{(3)}\wedge * F^{(3)}=0\,,\quad d*F^{(3)}- H^{(3)}\wedge F^{(5)}=0\,,\quad F^{(5)}=*F^{(5)}\,.
\label{eomB}
\eea

\subsection{T-duality}

T-duality in a direction $y$ adds a `leg' along $y$ to RR field components in directions orthogonal to $y$, thus transforming a $p$-form RR field into a $p+1$-form field. Hence $p$-forms with $p>3$ for IIA and $p>4$ for IIB could be generated. For this reason the action of T-duality is most simply expressed in a `democratic' formalism in which one extends the field content to include all the $p$-form fields with $p=1,3,5,7$ for IIA and $p=0,2,4,6,8$ for IIB. The higher $p$-form fields are defined in terms of the lower ones by the duality relations
\be
*F^{(p)}= (-1)^{[{p\over 2}]}F^{(10-p)}\,,
\label{duality}
\ee
 where the field strengths $F^{(p)}$ are defined as in (\ref{fieldstrengths}).

On this extended field space T-duality acts as follows.
Let us denote by $y$ the direction along which one performs T-duality and by $x^\mu$ the remaining coordinates. It is convenient to write the metric, B-field and field strengths as
\bea
ds^2 &=& G_{yy} (dy+A_\mu dx^\mu)^2 + \hat{g}_{\mu\nu} dx^\mu dx^\nu\,,\nn
B^{(2)}&=& B_{\mu y} dx^\mu \wedge (dy+A_\mu dx^\mu) + \hat{B}^{(2)}\,,\nn
F^{(p)}&=&F_y^{(p-1)}\wedge  (dy+A_\mu dx^\mu) + \hat{F}^{(p)}\,,
\label{trule0}
\eea
where the forms $\hat{B}^{(2)}$, $F_y^{(p-1)}$ and $\hat{F}^{(p)}$ are along the $x^\mu$ directions.

The T-duality transformed fields are
\bea
d{\tilde s}^2&=& G^{-1}_{yy} (dy+B_{\mu y} dx^\mu)^2 + \hat{g}_{\mu\nu} dx^\mu dx^\nu\,,\quad e^{2 \tilde \Phi}={e^{2\Phi}\over G_{yy}}\,,\nn
{\tilde B}^{(2)}&=& A_{\mu} dx^\mu dy + \hat{B}^{(2)}\,,\nn
{\tilde F}^{(p)}&=& \hat{F}^{(p-1)}\wedge(dy+B_{\mu y} dx^\mu)+F_y^{(p)}\,.
\label{trule}
\eea

Let us  summarize the steps needed to perform T-duality: 

1) Start from a configuration in the `physical' formalism, i.e., having field strengths with $p\le 5$, with fields satisfying equations (\ref{bianchi}) and (\ref{eomA}) or (\ref{eomB}).

2) Go to the `democratic'  formalism, by adding the higher order field strengths satisfying the duality relations (\ref{duality});

3) Perform T-duality, with the rules (\ref{trule}); the duality transformed fields will automatically preserve the duality constraints  (\ref{duality});

4) Return to the `physical' formalism if desired by dropping the the field strengths with $p>5$; the resulting fields will again satisfy the equations  (\ref{bianchi}) and  (\ref{eomA}) or (\ref{eomB}).

\subsection{S-duality}

To perform the S-duality in type II B string theory, one works in the `physical' formalism, and defines
the complex field: 
\be
\tau = C^{(0)}+i e^{-\Phi}\,.
\ee
The S-duality transformed fields are:
\bea
&&d{\tilde s}^2= |\tau| ds^2\,,\quad \tilde \tau = -{1\over \tau}\,,\nn
&&\tilde B^{(2)}= C^{(2)}\,,\quad {\tilde C}^{(2)}= -B^{(2)}\,,\quad {\tilde F}^{(5)}= F^{(5)}\,.
\eea

\end{document}